\documentclass[twocolumn, 10pt, showpacs, prl]{revtex4-1}

\usepackage[usenames,dvipsnames,svgnames,table]{xcolor}

\usepackage{amsmath}
\usepackage{amssymb}
\usepackage{amsthm}
\usepackage{graphicx}
\usepackage{epstopdf}

\usepackage{amsmath}
\usepackage{amssymb}
\usepackage{amsthm}
\usepackage{graphicx}
\usepackage{caption}

\newcommand{\beq}{\begin{equation}}
\newcommand{\eeq}{\end{equation}}

\newcommand{\ba}{\begin{array}}
\newcommand{\ea}{\end{array}}

\newcommand{\bea}{\begin{eqnarray}}
\newcommand{\eea}{\end{eqnarray}}

\newcommand{\bc}{\begin{center}}
\newcommand{\ec}{\end{center}}

\newcommand{\bt}{\begin{tabular}}
\newcommand{\et}{\end{tabular}}

\def\bi{\begin{itemize}}
\newcommand{\ei}{\end{itemize}}

\newcommand{\bd}{\begin{description}}
\newcommand{\ed}{\end{description}}

\newcommand{\bp}{\begin{pmatrix}}
\newcommand{\ep}{\end{pmatrix}}

\newcommand{\p}{\partial}
\newcommand{\sech}{\mbox{sech}}

\newcommand{\bv}{{\bf v}}

\def\br{{\bf r}}

\newcommand{\bdd}{{\bf d}}

\newcommand{\eq}{\mbox{Eq.}}

\def\digamma{S}

\def\boldc{\mathbf{c}}

\begin{document}
\title{Strong transmission and reflection of edge modes in bounded photonic graphene}

\author{Mark J. Ablowitz$^1$, Yi-Ping Ma$^{1,}$}
\email{yiping.m@gmail.com}
\affiliation{$^1$Department of Applied Mathematics, University of Colorado, Boulder, Colorado 80309, USA}

\begin{abstract}
The propagation of linear and nonlinear edge modes in bounded photonic honeycomb lattices formed by an array of rapidly varying helical waveguides is studied. These edge modes are found to exhibit strong transmission (reflection) around sharp corners when the dispersion relation is topologically nontrivial (trivial), and can also remain stationary.
An asymptotic theory is developed that establishes the presence (absence) of edge states on all four sides, including in particular armchair edge states, in the topologically nontrivial (trivial) case.
In the presence of topological protection,
nonlinear edge solitons can
persist over very long distances.
\end{abstract}
\maketitle

There is significant interest in using optical materials
to realize
phenomena that are difficult to observe in traditional materials. In particular,
light propagation in photonic lattices with a honeycomb (HC)  background, referred to as photonic graphene due to its similarities to material graphene~\cite{geim}, has been extensively studied both experimentally~\cite{Segev1,Segev4} and theoretically~\cite{abl1,feffwein}. In infinite, or bulk, lattices the wave dynamics exhibit the interesting feature of conical diffraction~\cite{Segev1}. This phenomenon has been explained by the presence of Dirac points, or conical intersections between dispersion bands \cite{abl3}.

Recently it was  shown  that introducing edges and suitable waveguides in the direction of propagation, unidirectional edge wave propagation occurs at optical frequencies \cite{rechts3}. The system is described analytically by the normalized lattice nonlinear Schr\"{o}dinger (NLS) equation
\begin{equation}
i\partial_{z}\psi = -(\nabla + i {\bf A}(z))^{2}\psi +V(\br)\psi - \sigma_0 \left|\psi \right|^{2} \psi,
\label{LNLS}
\end{equation}
where the scalar field $\psi$ is the complex envelope of the electric field, $z$ is the direction of propagation and takes on the role of time, $\br=(x,y)$ is the transverse plane, $\nabla\equiv(\partial_x,\partial_y)$, the potential $V(\br)$ is taken to be of HC type, and the coefficient $\sigma_0$ is the strength of the nonlinear change in the index of refraction. The vector field ${\bf A}(z)$ is determined by helical variation of the HC lattice
in the direction of propagation, and plays the role of a pseudo-magnetic field. The particular choice used in \cite{rechts3} is, in terms of dimensionless  coordinates,
\begin{equation}\label{eq:A-rechts}
{\bf A}(z)=\kappa (\sin{\Omega z}, -\cos{\Omega z}),
\end{equation}
where $\kappa$ and $\Omega$ are constant.
The waveguides defined by
${\bf A}(z)$ are
written into the optical lattice using a femtosecond laser writing technique~\cite{szameit2010}. The associated linear edge wave propagation
was investigated  experimentally and computationally in the tight binding limit in \cite{rechts3}. Remarkably, these edge waves were found to be nearly immune to backscattering. This phenomenon was found to be related to symmetry breaking perturbations which  separate the Dirac points and leave a nontrivial integer ``topological'' charge on the separated dispersion bands. In this setting, photonic graphene exhibits the hallmarks of Floquet topological insulators~\cite{lindner}.

Motivated by this work, the existence of traveling edge modes was investigated analytically in \cite{ACM14fast} for general periodic pseudo-fields ${\bf A}(z)$ and
with nonlinearity, i.e.~$\sigma_0\neq0$ in the tight binding limit. Assuming that the pseudo-field varies relatively rapidly, which is consistent with the experiments in \cite{rechts3}, an asymptotic theory based on Floquet theory
was developed which leads to a detailed description of the wave dynamics.
Importantly, in the presence of weak nonlinearity the classical one-dimensional NLS equation was found to govern the envelope of traveling edge modes.
When the NLS equation is focusing a family of nonlinear edge solitons exist due to the balance between dispersion and nonlinearity. When the dispersion relation is topologically nontrivial, these nonlinear edge solitons are immune to backscattering as with linear traveling edge modes.
The existence of these nonlinear edge solitons vastly expands the landscape and understanding of localized states along edges.

\begin{figure}
\centering \includegraphics[width=0.4\textwidth]{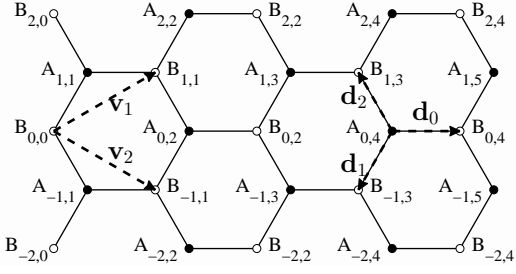}
\caption{A $5\times5$ bounded HC lattice with zig-zag edges on the left and right, and armchair edges on the top and bottom. The $A$ and $B$ sites are indexed following Ref.~\cite{abl5}. The primitive lattice vectors $\bv_1$ and $\bv_2$ and the intersite distances $\bdd_j$, $j=0,1,2$ are also labeled.} \label{fig:EDGE_bounded_hc_lattice}
\end{figure}

Much of the existing literature on edge modes in HC lattices, including the above referenced work, has focused on extended edges of either zig-zag or bearded type.  Here we consider a rectangular-type domain,
which has zig-zag edges on the left and right,
and armchair edges on the top and bottom, as illustrated in Fig.~\ref{fig:EDGE_bounded_hc_lattice}.
In the absence of helical waveguides, armchair edge states do not exist in isotropic HC lattices~\cite{KoHa07}.
However, with rapidly varying helical waveguides we find
traveling edge modes can exist along the armchair edges.
There are two types of armchair edge states corresponding respectively to isotropic and anisotropic HC lattices.
In the presence (absence) of these armchair edge states, linear edge modes and nonlinear edge solitons are found to exhibit strong transmission (reflection) around the sharp corners under certain conditions; otherwise they may decay due to dispersion or scattering.
The ability to easily switch between transmission and reflection  as well as remaining stationary, by varying the
underlying parameters provides a new means for the control of light conferred by appropriately merging topological, linear and nonlinear effects.

To employ the tight binding approximation, we substitute  $\psi = e^{-i\br \cdot {\bf A}(z)}\phi$ into $\eq$ \eqref{LNLS},
and assume a Bloch wave envelope of the form  \cite{abl1}
\begin{equation}\label{eq:tb-ansatz}
\phi \sim \sum_{\bv }\left(a_{\bv}(z)\phi_{1,\bv} + b_{\bv}(z)\phi_{2,\bv} \right)
\end{equation}
where $\phi_{1,\bv}=\phi_{1}(\br-\bv)$, $\phi_{2,\bv}=\phi_{2}(\br-\bv)$ are the linearly independent orbitals associated with the two sites A and B where the HC potential $V(\br)$ has minima in each fundamental cell.
The lattice sites are located at $\bv=l_1\bv_{1}+l_2\bv_{2}$, where $l_1,l_2\in\mathbb{Z}$, and the lattice vectors $\bv_{1}$ and $\bv_{2}$ are given by $\bv_{1} = \left( \sqrt{3}/2, ~ 1/2\right), ~ \bv_{2} = \left(\sqrt{3}/2, ~ -1/2\right).$
Carrying out the requisite calculations (see \cite{abl1} for more details),
we arrive at the following two dimensional discrete system
\begin{align}
i\p_{z}a_{mn}  + (\mathcal{L}_{-}(z)b)_{mn} + \sigma|a_{mn}|^{2}a_{mn}= 0, \label{eq:abmn-a}\\
i\p_{z}b_{mn}  + (\mathcal{L}_{+}(z)a)_{mn}  + \sigma|b_{mn}|^{2}b_{mn}= 0, \label{eq:abmn-b}
\end{align}
where $\sigma$ depends on the underlying orbitals, is proportional to %
$\sigma_0$,
and the linear operators $\mathcal{L}_\pm$ take the form
\begin{align*}
(\mathcal{L}_{-}b)_{mn} = &e^{i\theta_0} b_{mn} +\rho(e^{i\theta_1}b_{m-1,n-1} +e^{i\theta_2}b_{m+1,n-1}), \\
(\mathcal{L}_{+}a)_{mn} =  &e^{-i\theta_0} a_{mn} +\rho(e^{-i\theta_1}a_{m+1,n+1} +e^{-i\theta_2}a_{m-1,n+1}),
\end{align*}
where $\rho$ is a lattice deformation parameter, and $\theta_j=\bdd_j\cdot{\bf A}, ~ j=0,1,2.$
The vector $\bdd_0$ is the direction vector from the site $B_{mn}$ to the site $A_{mn}$, and is approximately
$(1/\sqrt{3},0)$ in the tight binding limit~\cite{ACM14fast}. The vectors $\bdd_{1,2}=\bdd_0-\bv_{1,2}$ are the direction vectors from any given $B$ site to the other two neighboring $A$ sites.
For later reference we express the full 2D discrete system (\ref{eq:abmn-a}--\ref{eq:abmn-b}) in the compact form
\begin{equation}\label{eq:abmn-c}
i\p_zc_{mn}+(\mathcal{L}c)_{mn}+\sigma|c_{mn}|^2c_{mn}=0,
\end{equation}
where
\begin{equation*}
c=\left(\begin{array}{c}a\\b\end{array}\right),\quad
\mathcal{L}=\bp 0 & \mathcal{L}_{-}(z) \\ \mathcal{L}_{+}(z) & 0 \ep.
\end{equation*}
The HC lattice is formed by those sites with $m+n$ even as shown in Fig.~\ref{fig:EDGE_bounded_hc_lattice}.

\begin{figure}
\centering
\begin{tabular}{cc}
\includegraphics[width=0.24\textwidth]{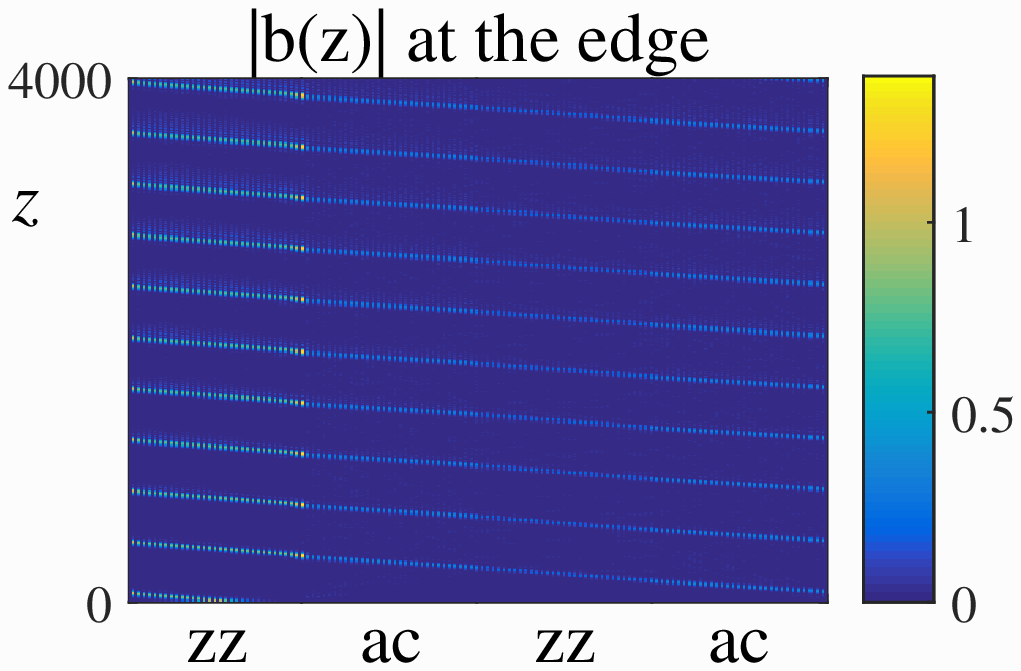} &
\includegraphics[width=0.24\textwidth]{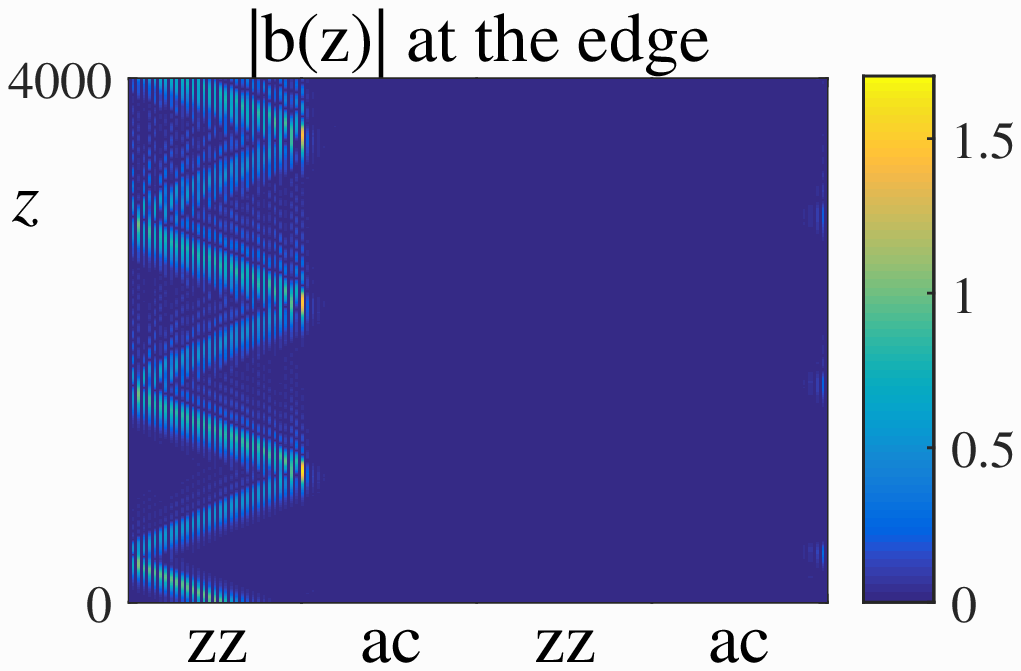} \\
(a) & (b)
\end{tabular}
\caption{Transmission and reflection of linear edge modes ($\sigma=0$) in a $65\times65$ bounded HC lattice, shown as space-time plots of $|b(z)|$ at the edges. Here `zz' represents zig-zag and `ac' represents armchair; the horizontal axis begins with the left zig-zag edge with increasing $m$. The pseudo-field is given by Eq.~(\ref{eq:A-rechts}) with $\kappa=1.4$ and period $T=1.5$, and the deformation parameter $\rho$ is (a) $\rho=1$; (b) $\rho=0.4$. The initial condition (\ref{eq:edso-ic-env}-\ref{eq:edso-ic-sech}) has $\omega_0=\pi/2$, $\nu=0.2$, and $A=1$. See~\cite{MovieTr} and~\cite{MovieRe} for animations of panels (a) and (b).} \label{fig:EDGE_zzac_line}
\end{figure}

Using the pseudo-field given by Eq.~(\ref{eq:A-rechts}) with period $T=2\pi/\Omega=1.5$  (consistent with the experiments in~\cite{rechts3}), our 2D discrete system (\ref{eq:abmn-c}) has two  free parameters $\rho$ and $\sigma$;
$\rho=1$ and $\sigma=0$ for a perfect HC lattice without nonlinearity in~\cite{rechts3}.
The initial condition is taken to be localized on the left zig-zag edge of the form
\begin{equation}\label{eq:edso-ic-env}
c_{mn}(z=0)=C(m,z=0)e^{i m\omega_0}c_n^E,
\end{equation}
where $\omega_0$ is the carrier wavenumber in $m$ (along the edge), and $c_n^E$ is the 1D edge state that decays in $n$ (perpendicular to the edge).
The envelope function $C$, whose evolution is determined below, is taken to be
\begin{equation}\label{eq:edso-ic-sech}
C(m,z=0)=A\,\sech(\nu (m-m_0)),
\end{equation}
where $A$ is the amplitude, $\nu$ is the spectral width
$,0<\nu \ll 1,$  and $m_0=0$
unless otherwise stated. The 2D evolution can be visualized using space-time plots of the amplitude $|b(z)|$ along the four edges, as shown in Fig.~\ref{fig:EDGE_zzac_line}. The effective mass of the mode for any $z$ is the square of the $L^2$-norm in an interval of width $4/\nu$ around the center of mass evaluated at the edge.

First we consider the linear case ($\sigma=0$), and choose the initial condition to have $\omega_0=\pi/2$, $\nu=0.2$,
and $A=1$ consistent with~\cite{rechts3}. In this case  we find that two contrasting types of  linear wave dynamics can exist in the bounded HC lattice depending on the choice of $\rho$. As shown in Fig.~\ref{fig:EDGE_zzac_line}(a), for $\rho=1>1/2$, the edge mode is almost perfectly transmitted around the sharp corners and thus loops around the HC lattice. The effective mass remaining after four (eight) loops, or $z=1550$ ($z=3100$), is $96\%$ ($94\%$). As shown in Fig.~\ref{fig:EDGE_zzac_line}(b), for $\rho=0.4<1/2$, the edge mode is very strongly reflected at the sharp corners and thus bounces back and forth along the left edge. The effective mass remaining after two (four) reflections, or $z=1250$ ($z=2500$), is $96\%$ ($89\%$). See \cite{MovieTr} and~\cite{MovieRe} for animations of Fig.~\ref{fig:EDGE_zzac_line}(a) and Fig.~\ref{fig:EDGE_zzac_line}(b).
For either choice of $\rho$, choosing $\omega_0\neq\pi/2$ causes the edge mode to disperse more and thus destroys coherence. Numerical results not shown here indicate that the overall wave dynamics are insensitive to the detailed shapes of the corners, though the edge mode lingers longer at the corners containing degree-1 vertices (e.g. the upper and lower left corners in Fig.~\ref{fig:EDGE_bounded_hc_lattice}).

To describe a zig-zag (armchair) edge mode
we take a discrete Fourier transform in $m$ ($n$) on Eq.~(\ref{eq:abmn-c}), i.e.~letting $c_{mn}=c_{n}e^{im\omega_m}$ ($c_{mn}=c_{m}e^{in\omega_n}$).
Following \cite{rechts3, ACM14fast}, we assume that the pseudo-field ${\bf A}$ or equivalently the linear operator $\mathcal{L}$ varies rapidly, namely $\mathcal{L}=\mathcal{L}(\zeta)$, where $\zeta = \epsilon^{-1} z,  |\epsilon| \ll1$, and thus has period $\tilde{T}=\epsilon^{-1}T$.  Then we can employ a multiple-scales ansatz $\boldc = \boldc(z,\zeta)$,
and expand $\boldc$ in powers of $\epsilon$ as $\boldc = \boldc^{(0)} + \epsilon \boldc^{(1)} + \cdots$, where $\boldc$ denotes the vector $c_{mn}$, $c_n$ or $c_m$ depending on the context.
At $O(\epsilon^{-1})$, $\partial_\zeta \boldc^{(0)}=0$ which leads to $\boldc^{(0)}=\boldc^{(0)}(z)$. At $O(1)$, to remove secularities, we get
\begin{equation}\label{eq:c_z_O1}
i\partial_z\boldc^{(0)}+\bar{\mathcal{L}}\boldc^{(0)}+\sigma|\boldc^{(0)}|^2\boldc^{(0)}=0,
\end{equation}
where $\bar{\mathcal{L}}$ is given by the Magnus-type  expansion~\cite{BCOR09}
\begin{equation}\label{eq:r_pm}
\bar{\mathcal{L}}=\bar{\mathcal{L}}^{(0)}+\epsilon\bar{\mathcal{L}}^{(1)}+\epsilon^2\bar{\mathcal{L}}^{(2)}+O(\epsilon^3),
\end{equation}
with $\bar{\mathcal{L}}^{(0)}=\tilde{T}^{-1}\int_0^{\tilde{T}}\mathcal{L}(\zeta)d\zeta,$
\begin{equation*}
\bar{\mathcal{L}}^{(1)}=\frac{i}{2\tilde{T}}\int_0^{\tilde{T}}\int_0^\zeta(\mathcal{L}(\zeta)\mathcal{L}(\zeta')-\mathcal{L}(\zeta')\mathcal{L}(\zeta))d\zeta'd\zeta
\end{equation*}
and $\bar{\mathcal{L}}^{(2)}$ can be written in terms of a triple integral not explicitly shown. The discrete system (\ref{eq:c_z_O1}) conserves both
the mass $N$ and energy $H$ given by
\begin{align}
&N=\langle\boldc^{(0)},\boldc^{(0)}\rangle,\label{eq:conserve-mass}\\
&H=-\langle\boldc^{(0)},\bar{\mathcal{L}}\boldc^{(0)}\rangle-\langle\boldc^{(0)},\sigma|\boldc^{(0)}|^2\boldc^{(0)}\rangle/2,\label{eq:conserve-energy}
\end{align}
where the inner product is defined as $\langle\mathbf{f},\mathbf{g}\rangle\equiv\sum_jf_j^*g_j$.

Assuming that the spectral envelope is centered at $\omega_0$ with width $\nu$, $0<\nu\ll1$, the edge mode may be expressed in terms of a scalar envelope function $C(z;\omega)$ as
\begin{equation}
\boldc^{(0)}=C(z;\omega)\boldc^E
\end{equation}
where $\boldc^E$ is the 1D edge state given by
\begin{equation}\label{eq:lin-edge-gen}
\bar{\mathcal{L}}(\omega)\boldc^E=-\alpha(\omega)\boldc^E,
\end{equation}
normalized such that $\|\boldc^E\|_2^2= \sum_{j}\left|c_{j}^{E} \right|^{2}=1$. The function $\alpha(\omega)$ is the dispersion relation.
The solvability condition on Eq.~(\ref{eq:c_z_O1})
 leads to
\begin{equation}\label{eq:env-eq-gen}
i\partial_zC-\alpha(\omega)C+\tilde{\sigma}(\omega)|C|^2C=0,
\end{equation}
where $\tilde{\sigma}(\omega)\equiv\sigma\alpha_{nl}(\omega)$ with $\alpha_{nl}(\omega) = \|\boldc^{E}\|_{4}^{4}= \sum_{j}\left| c_{j}^{E} \right|^{4}$.

When the nonlinearity $\sigma \neq 0$ and second order dispersion $\alpha''(\omega_0)\neq0$, maximal balance between dispersion and nonlinearity leads to
\begin{equation}\label{eq:env-eq-clnls}
i\partial_zC-\left(\alpha_0+\alpha_0'\tilde{\omega}+\frac{1}{2}\alpha_0''\tilde{\omega}^2\right)C+\tilde{\sigma}_0|C|^2C=0,
\end{equation}
where $\tilde{\omega}\equiv\omega-\omega_0$, $\alpha_0^{(j)}\equiv\alpha^{(j)}(\omega_0)$, $j=0,1,2$, and $\tilde{\sigma}_0\equiv\tilde{\sigma}(\omega_0)$. The group velocity of the envelope is $\alpha_0'$. We remark that the special stationary case can occur when $\alpha_0'=0$; in general if the pseudo-field ${\bf A}$ is independent of $z$, then the time-independent linear operator $\mathcal{L}$ can admit the so-called ``zero-energy'' edge states with $\alpha(\omega)=0$ identically~\cite{KoHa07}. Replacing $\tilde{\omega}$ by $-i\partial_{\chi}$ where $\chi$ represents the $m$ ($n$) direction for the zig-zag (armchair) edge, Eq.~(\ref{eq:env-eq-clnls}) becomes the 1D classical second-order focusing (defocusing) NLS equation when $\alpha_0''\sigma>0$ ($\alpha_0''\sigma<0$). In the focusing case, Eq.~(\ref{eq:env-eq-clnls}) admits a two-parameter family of solitons~\cite{MJA2011};
the two parameters may be chosen as $\omega_0$ and $\nu$.
A 2D edge soliton  is then obtained from the initial conditions (\ref{eq:edso-ic-env}--\ref{eq:edso-ic-sech}) with
\begin{equation} \label{amplitude}
A=\nu\sqrt{\alpha_0''/\tilde{\sigma}_0}.
\end{equation}
Thus robust edge solitons exist for a finite interval of $\omega_0$;
at each $\omega_0$,
the choice of $\nu$ fixes the amplitude $A$.

\begin{figure}
\centering
\begin{tabular}{cc}
\includegraphics[width=0.24\textwidth]{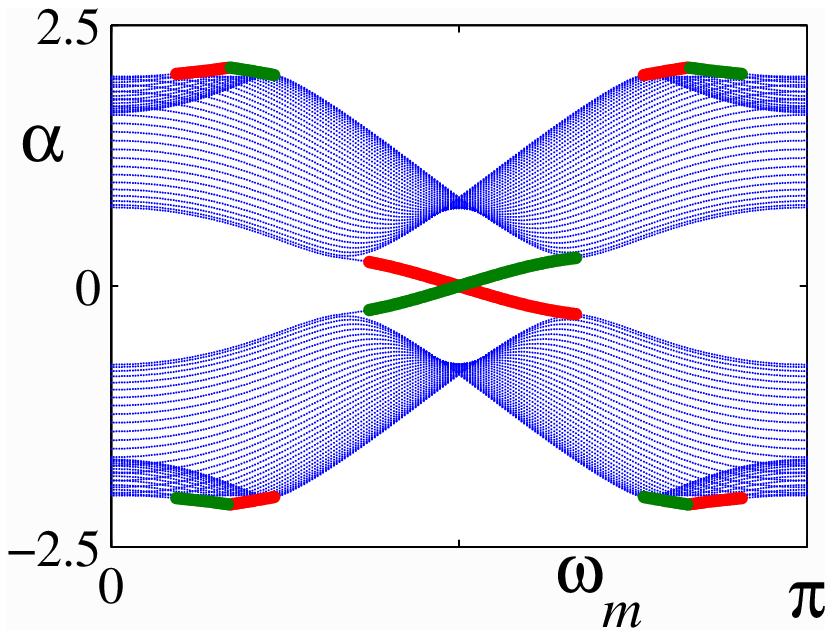} &
\includegraphics[width=0.24\textwidth]{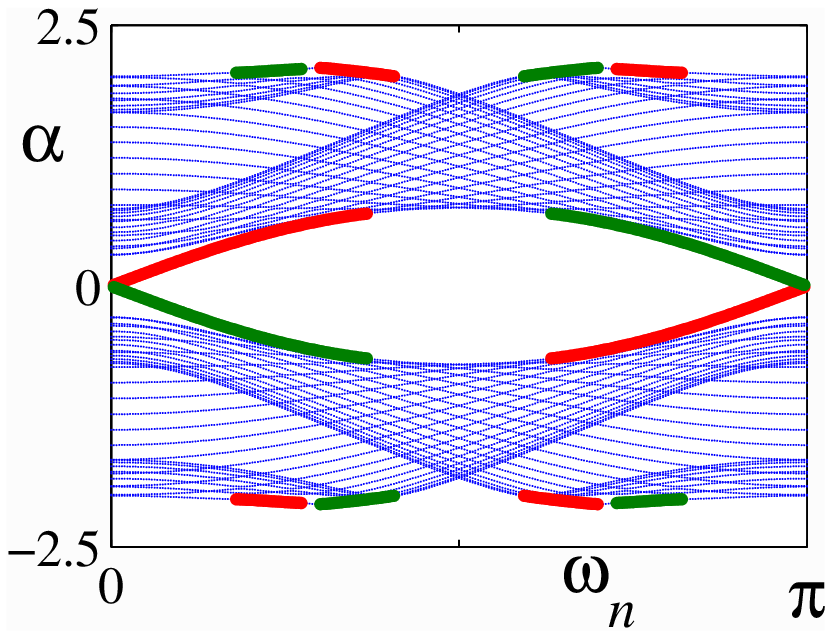} \\
(a) & (b) \\
\includegraphics[width=0.24\textwidth]{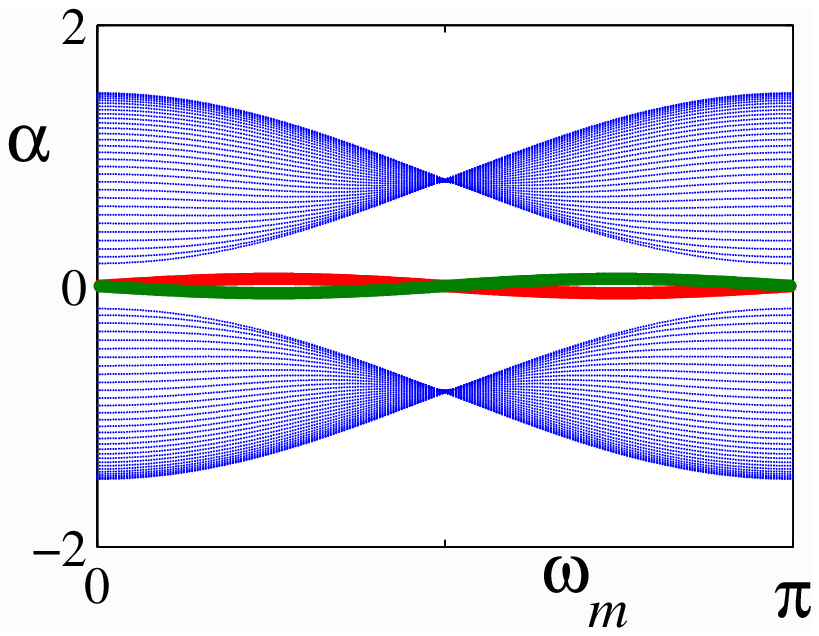} &
\includegraphics[width=0.24\textwidth]{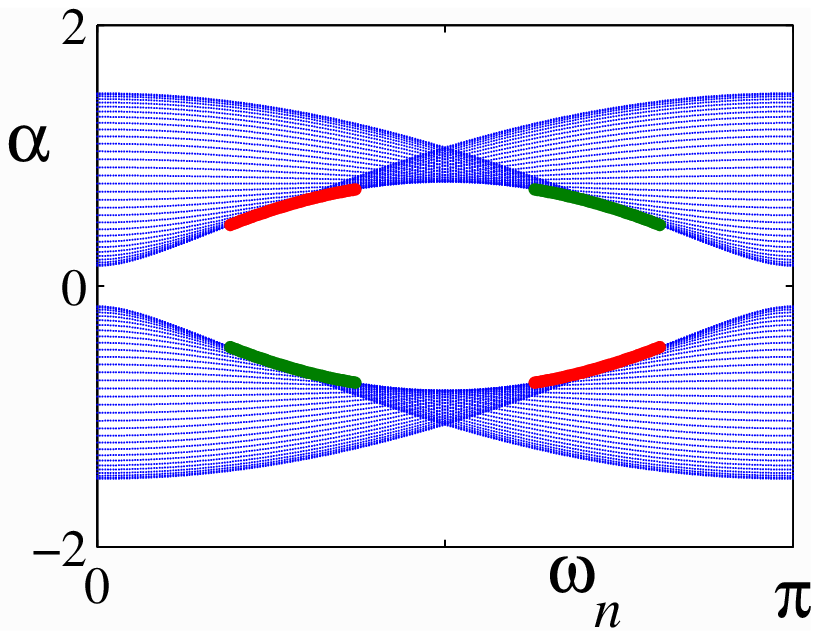} \\
(c) & (d)
\end{tabular}
\caption{Dispersion relations on the (a,c) zig-zag and (b,d) armchair edges.
Values of ${\bf A}$ and $\rho$
in panels (a,b) and (c,d)  agree respectively with Fig.~\ref{fig:EDGE_zzac_line}(a) and Fig.~\ref{fig:EDGE_zzac_line}(b). }

\label{fig:EDGE_zzac_disp}
\end{figure}

The zig-zag (armchair) dispersion relation can be numerically computed  on a 1D domain with suitable  boundary conditions on both ends and compared with analytic predictions based on Eq.~(\ref{eq:lin-edge-gen}). Figure~\ref{fig:EDGE_zzac_disp} shows the dispersion relations computed using the same parameters as in Fig.~\ref{fig:EDGE_zzac_line} and plotted on the periodic interval $\omega\in[0,\pi)$. The Floquet eigenvalues corresponding to the eigenfunctions with at least  90\% of the mass concentrated on the left (right) half domain are highlighted in red (green). The $\mathbb{Z}_2$ topological index $I$ is defined as the number of intersections modulo 2 between either the red or green curve and the horizontal axis~\cite{ACM14fast}. Interestingly, in some cases there exist edge states near the outer boundaries of the bulk spectrum at large $|\alpha|$, but since they are not related to the wave dynamics at small $|\alpha|$, they will not be discussed further here.

As established in~\cite{ACM14fast} for the zig-zag edge, to leading order the 1D edge states $\boldc^E$ satisfy
\begin{equation}\label{eq:lin-edge-zz-mode}
\bar{\mathcal{L}}^{(0)}\boldc^E=0.
\end{equation}
For those edge states localized on the left (right), the mass is concentrated in the $b$ ($a$) sites. The decay exponents in $n$ are $O(1)$,
with amplitudes $O(1)$ on the edge. The dispersion relation $\alpha(\omega)$ is $O(\epsilon)$ and is given by
\begin{equation}\label{eq:lin-edge-zz-disp}
\alpha(\omega)=\epsilon\langle\boldc^E,\bar{\mathcal{L}}^{(1)}\boldc^E\rangle/\langle\boldc^E,\boldc^E\rangle;
\end{equation}
thus the group velocity of the envelope is $O(\epsilon)$. The zig-zag dispersion relation computed using the same parameters as in Fig.~\ref{fig:EDGE_zzac_line}(a) (Fig.~\ref{fig:EDGE_zzac_line}(b)) is shown in Fig.~\ref{fig:EDGE_zzac_disp}(a) (Fig.~\ref{fig:EDGE_zzac_disp}(c)). This dispersion relation is topologically nontrivial ($I=1$) for $\rho=1>1/2$ and topologically trivial ($I=0$) for $\rho=0.4<1/2$. In the former case, the left zig-zag dispersion relation terminates at two points denoted respectively by $(\omega_{m-},\alpha_-)$ near the upper bulk and $(\omega_{m+},-\alpha_+)$ near the lower bulk, while the right zig-zag dispersion relation has the opposite sign. These four points, known as the massive Dirac points, result from splitting the massless Dirac points in $\bar{\mathcal{L}}^{(0)}$.

For the armchair edge, $\bar{\mathcal{L}}^{(0)}$ describes isotropic graphene when ${\bf A}$ is invariant under $2\pi/3$ rotations, and anisotropic graphene otherwise. Thus the circular pseudo-field given by Eq.~(\ref{eq:A-rechts}) leads to isotropic graphene. Anisotropic graphene may be exemplified using the elliptic pseudo-fields invariant under $\pi$ rotations introduced in \cite{ACM14fast}.
In this case
the edge modes behave similarly to  those in the zig-zag case.

In the isotropic armchair case, $\bar{\mathcal{L}}^{(0)}$ does not admit edge states~\cite{KoHa07}.
However, as shown in Fig.~\ref{fig:EDGE_zzac_disp}(b) and Fig.~\ref{fig:EDGE_zzac_disp}(d) computed using the same parameters as in Fig.~\ref{fig:EDGE_zzac_line}(a) and Fig.~\ref{fig:EDGE_zzac_line}(b), armchair edge states exist near the inner boundaries of the bulk spectrum. The topological index of the armchair dispersion relation is found to be the same as
the zig-zag case, namely $I=1$ for $\rho>1/2$ and $I=0$ for $\rho<1/2$. The existence of these edge states in $\bar{\mathcal{L}}$ can be explained by expanding $\bar{\mathcal{L}}$ to $O(\epsilon^2)$ in Eq.~(\ref{eq:lin-edge-gen}).
For these armchair edge states, the mass is equally distributed in the $a$ and $b$ sites. The decay exponents in $m$ are $O(\epsilon)$, with amplitudes $O(\epsilon^{1/2})$ on the edge.
The dispersion relation $\alpha(\omega)$ is $O(1)$, and so the group velocity of the envelope is also $O(1)$. The detailed calculations will be presented in a separate paper. Here we note that in the $I=1$ case as shown in Fig.~\ref{fig:EDGE_zzac_disp}(b), the armchair dispersion relation does not terminate at the massive Dirac points, which are located at $(0,\pm\alpha_-)$ and $(0,\pm\alpha_+)$ and thus are degenerate in $\omega_n$.

The linear and nonlinear dynamics of edge waves can
be understood in terms of the dispersion relations and the conservation of mass $N$ and  energy $H$ defined in  Eqs.~(\ref{eq:conserve-mass}--\ref{eq:conserve-energy}).
Consider a $J$-soliton configuration where each soliton is given by Eq.~(\ref{eq:edso-ic-env}) with envelope function $C_j$, carrier wavenumber $\omega_j$ and carrier frequency $\alpha_j$, $j=1,\cdots,J$. Then to leading order, the mass and energy evaluate to
\begin{equation}
N=\sum_{j=1}^J\|C_j\|_2^2, \quad H=-\sum_{j=1}^J\alpha_j\|C_j\|_2^2,
\end{equation}
where $\|\cdot\|_2$ denotes the $L^2$-norm.

In the linear case, the amplitude $|C|$ and frequency $\alpha$ of each individual wavenumber $\omega$ is conserved. Let $\mathcal{I}_\alpha$ denote the interval in $\alpha$ spanned by the left zig-zag dispersion relation. When $\rho>1/2$, both the zig-zag and the armchair dispersion relations are topologically nontrivial as shown in Figs.~\ref{fig:EDGE_zzac_disp}(a--b). In this case, corresponding to any $\alpha\in\mathcal{I}_\alpha$ there is a single value of $\omega$ and thus a single unidirectional edge mode on each of the four edges. The
strong transmission of linear edge modes in Fig.~\ref{fig:EDGE_zzac_line}(a) is then explained by a global counterclockwise mode formed by `gluing' together these four edge modes through the four corners. When $\rho<1/2$, both the zig-zag and the armchair dispersion relations are topologically trivial as shown in Figs.~\ref{fig:EDGE_zzac_disp}(c--d). In this case, corresponding to any $\alpha\in\mathcal{I}_\alpha$ there are two values of $\omega$ and thus a pair of counter-propagating edge modes on either of the two zig-zag edges. The strong reflection of linear edge modes in Fig.~\ref{fig:EDGE_zzac_line}(b) is then explained by the pair of edge modes on the left zig-zag edge transforming into each other via scattering at the upper and lower left corners. In short, the transmission channel is available and the reflection channel is unavailable when the dispersion relation is topologically nontrivial, and vice versa. We remark that for other parameter choices, the transmission and reflection channels may be simultaneously available, as well as the scattering channel into the bulk. These possibilities results in weaker transmission/reflection, and so are outside the scope of this paper.

\begin{figure}
\centering
\begin{tabular}{cc}
\includegraphics[width=0.24\textwidth]{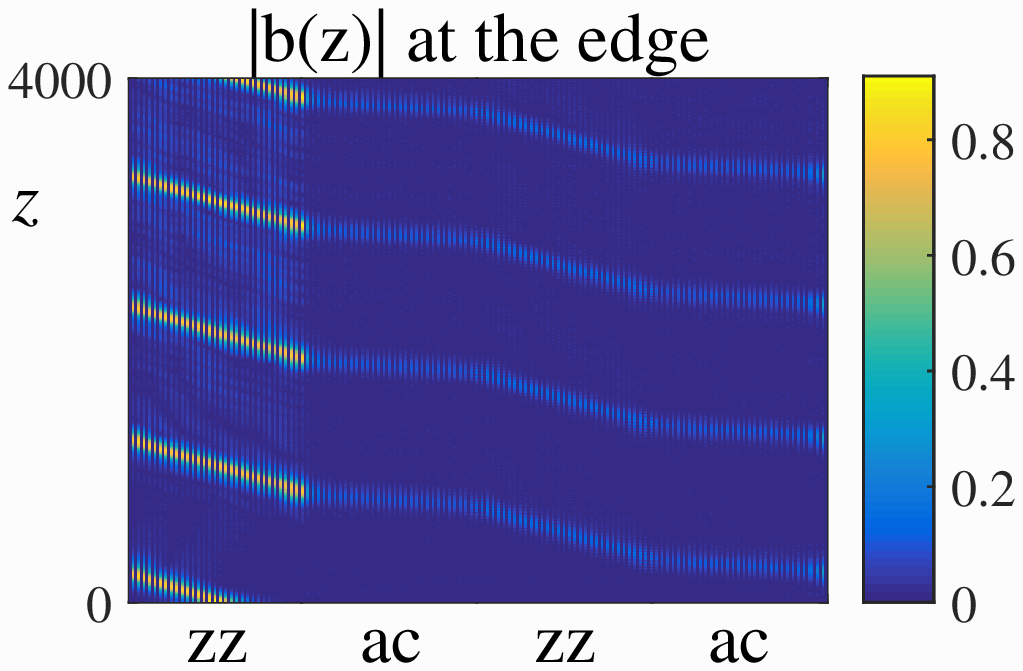} &
\includegraphics[width=0.24\textwidth]{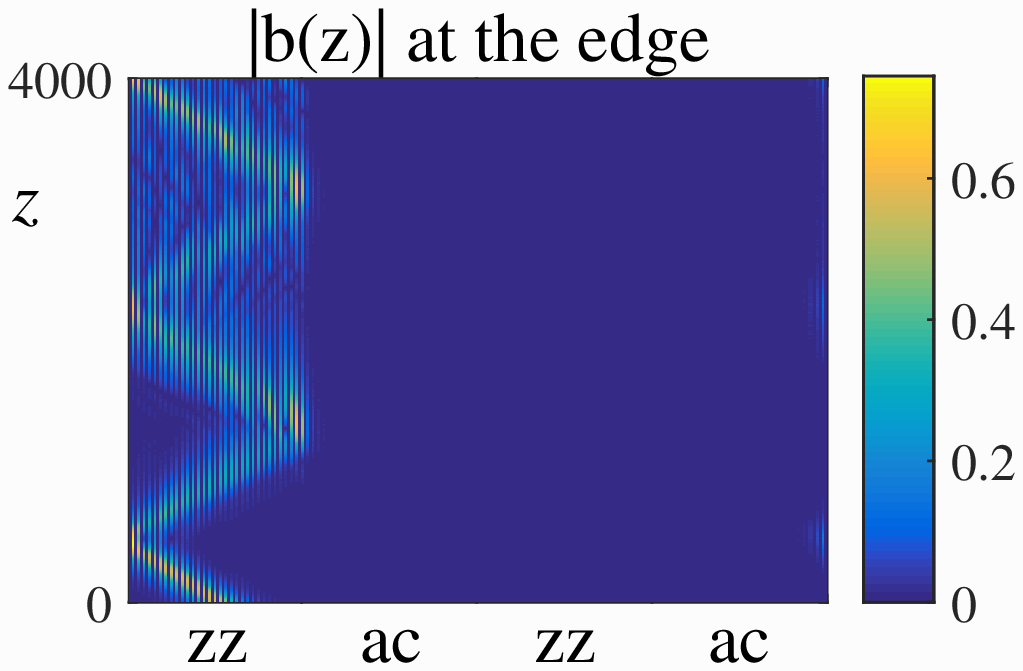} \\
(a) & (b) \\
\includegraphics[width=0.24\textwidth]{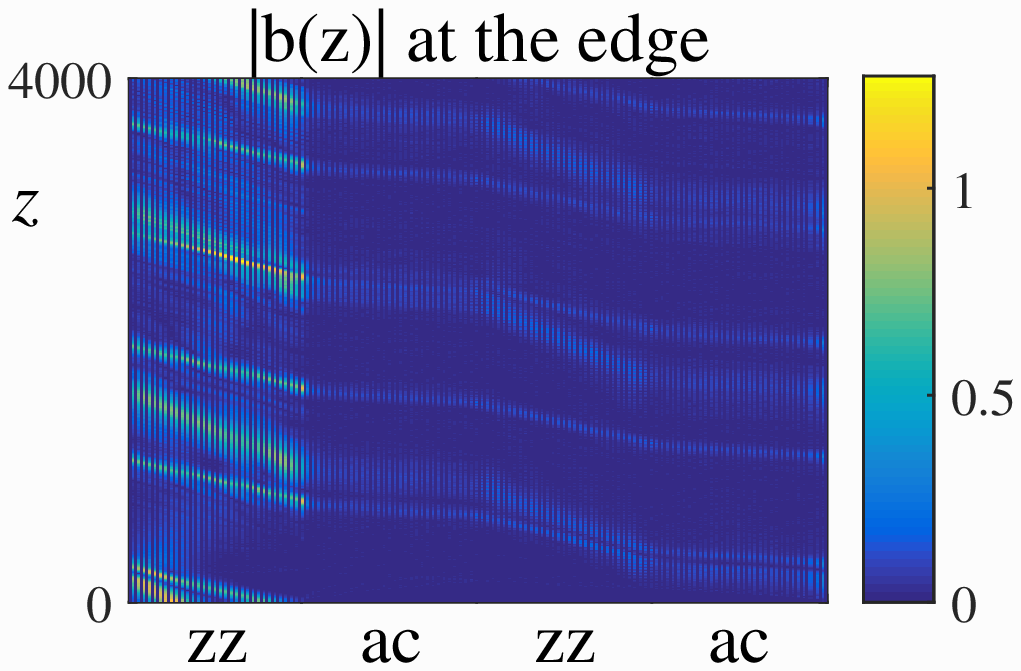} &
\includegraphics[width=0.24\textwidth]{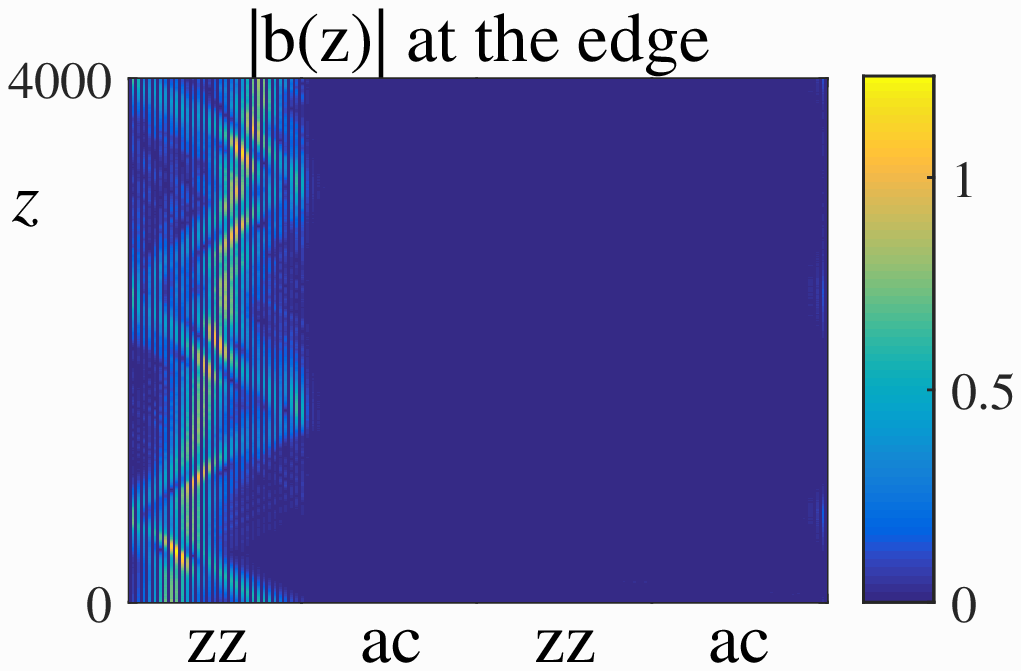} \\
(c) & (d)
\end{tabular}
\caption{Transmission and reflection of nonlinear edge solitons ($\sigma=0.02$) in a $65\times65$ bounded HC lattice, shown as space-time plots of $|b(z)|$ at the edges. Values of ${\bf A}$ and
$\rho$ in panels (a,c) and (b,d)
agree respectively with Fig.~\ref{fig:EDGE_zzac_line}(a) and Fig.~\ref{fig:EDGE_zzac_line}(b), except that $\rho=0.6$ in panels (a,c). The initial condition (\ref{eq:edso-ic-env}--\ref{eq:edso-ic-sech}) has $A$ given by Eq.~(\ref{amplitude}), $\nu=0.2$ and (a,b) $\omega_0=5\pi/8$.
Panels (c,d) have two envelopes with $(m_0,\omega_0)$: (c) $(-16,2\pi/3)$ \& $(16,7\pi/12)$; (d) $(-16,3\pi/4)$ \& $(16,5\pi/8)$.} \label{fig:EDGE_zzac_nlin}
\end{figure}

Next we consider a nonlinear case: $\sigma=0.02\neq0$, with initial conditions
$\omega_0=5\pi/8$, $\nu=0.2$, and $A$ given by Eq.~(\ref{amplitude}). In this case the 1D NLS equation (\ref{eq:env-eq-clnls}) is focusing and so admits edge solitons on the zig-zag edges. In order that the zig-zag dispersion relation spans a wider interval in $\omega_m$ and thus has greater curvature $\alpha''$, we choose smaller $\rho$ in the topologically nontrivial case following~\cite{ACM14fast}. As shown in Fig.~\ref{fig:EDGE_zzac_nlin}(a) for $\rho=0.6>1/2$,
the edge soliton is almost perfectly transmitted around the sharp corners. The effective mass remaining after two (four) loops, or $z=2000$ ($z=4000$), is $93\%$ ($93\%$). As shown in Fig.~\ref{fig:EDGE_zzac_nlin}(b) for $\rho=0.4<1/2$, the edge soliton is strongly reflected, but noticeably not as well as in the linear case, at the sharp corners. The effective mass remaining after two (four) reflections, or $z=1650$ ($z=3300$), is $72\%$ ($71\%$).

In a simple scattering process where an incoming edge soliton is scattered by one or more corners into an outgoing edge soliton, the conservation of the carrier frequency $\alpha$ determines the carrier wavenumber $\omega$ as in the linear case, and the conservation of the envelope mass $\|C\|_2^2$, which amounts to the conservation of $\nu\alpha_0''/\tilde{\sigma}_0$ using Eqs.~(\ref{eq:edso-ic-sech}) and (\ref{amplitude}), determines the spectral width $\nu$. In the transmission case ($\rho>1/2$), since an incoming zig-zag edge soliton is scattered by an even number of corners into an outgoing zig-zag edge soliton almost perfectly,
the conservation laws imply that the zig-zag edge soliton also maintains its shape almost perfectly. In the reflection case ($\rho<1/2$), the zig-zag edge soliton does not maintain its shape as well
due to multi-scattering at the corners and backscattering along the edge.

Finally we consider soliton interactions in the nonlinear case, with the initial condition chosen to contain two zig-zag edge solitons with different carrier wavenumbers $\omega_0$ and thus different group velocities. In the transmission (reflection) case, as shown in Fig.~\ref{fig:EDGE_zzac_nlin}(c) (Fig.~\ref{fig:EDGE_zzac_nlin}(d)), the two edge solitons are transmitted (reflected) almost independently. The collision between these two edge solitons
exhibits a phase shift, reminiscent of soliton collisions in the classical 1D NLS equation.

In this paper, strong transmission and reflection of linear and nonlinear edge modes is demonstrated in bounded photonic graphene formed by rapidly varying periodic waveguides. The transmission (reflection) phenomenon depends on the presence (absence) of topological protection on the zig-zag and armchair edges. For isotropic graphene the armchair edge states have slowly decaying profiles.
These unconventional edge states provide an alternative means to localize and transport light on armchair edges without
additional modifications~\cite{MaRoSb13}.
Simple scattering of edge modes by sharp corners is explained via conservation of mass and energy. In the presence of nonlinearity, the existence of nonlinear edge solitons greatly broadens the landscape of coherent edge modes. The synergy between topology and nonlinearity makes bounded photonic graphene an ideal candidate for robust routing of electromagnetic energy, which is a topic of considerable recent interest~\cite{MKMS15}.

This research was partially supported by the U.S. Air Force Office of Scientific Research, under grant FA9550-12-1-0207 and by the NSF under grants CHE 1125935 and DMS-1310200.

\end{document}